\documentclass[12pt]{article}
\usepackage{amssymb}
\usepackage{bbm}
\usepackage{epsfig}
\usepackage{array}
\usepackage{float}
\usepackage{dsfont}
\usepackage{amstext}
\usepackage{amsmath}
\usepackage{psfrag}
\usepackage{a4}
\usepackage{a4wide}
\usepackage{hyperref}
\usepackage{feynmp-auto}
\def\be{\begin{equation}}
\def\ee{\end{equation}}
\def\bea{\begin{eqnarray}}
\def\eea{\end{eqnarray}}

\begin{document}

\begin{center}
\baselineskip 20pt 
{\Large\bf Gravity Waves and Proton Decay in Flipped
SU(5) Hybrid Inflation Model}

\vspace{1cm}

{\large 
Mansoor Ur Rehman$^{a,}$\footnote{Email: \texttt{\href{mailto:mansoor@qau.edu.pk}{mansoor@qau.edu.pk}}}, Qaisar Shafi$^{b,}$\footnote{Email: \texttt{\href{mailto:shafi@bartol.udel.edu}{shafi@bartol.udel.edu}}} and Umer Zubair$^{b,}$\footnote{Email: \texttt{\href{mailto:uzubair@student.qau.edu.pk}{umer@udel.edu}}} 
} 
\vspace{.5cm}

{\baselineskip 20pt \it
$^a$Department of Physics,  \\
Quaid-i-Azam University, Islamabad 45320, Pakistan \\
\vspace*{6pt}
$^b$Bartol Research Institute, Department of Physics and Astronomy,  \\
University of Delaware, Newark, DE 19716, USA \\
\vspace{2mm} }

\vspace{1cm}
\end{center}

\begin{abstract}
We revisit supersymmetric hybrid inflation in the context of flipped $SU(5)$ model. With minimal superpotential and minimal K\"ahler potential, and soft SUSY masses of order $(1 - 100)$ TeV, compatibility with the Planck data yields a symmetry breaking scale $M$ of flipped $SU(5)$ close to $(2 - 4) \times 10^{15}$ GeV. This disagrees with the lower limit $M \gtrsim 7 \times 10^{15}$ GeV set from proton decay searches by the Super-Kamiokande collaboration. We show how $M$ close to the unification scale $2\times 10^{16}$ GeV  can be reconciled with SUSY hybrid inflation by employing a non-minimal K\"ahler potential.
Proton decays into $e^+ \pi^0$ with an estimated lifetime of order $10^{36}$ years. The tensor to scalar ratio $r$ in this case can approach observable values $\sim 10^{-4} - 10^{-3}$. 
\end{abstract}

%
\section{\large{\bf Introduction}}%
Supersymmetric (SUSY) hybrid inflation model~\cite{Dvali:1994ms,Copeland:1994vg,Linde:1997sj,Chamseddine:1982jx,Senoguz:2003zw} has attracted a fair amount of attention due to its simplicity and eleganance in realizing the grand unified theory (GUT) models of inflation \cite{Senoguz:2003zw}. In models with minimal K\"ahler potential, the soft linear and mass squared terms play important role in attaining the scalar spectral index compatible with the current experimental observations \cite{Rehman:2009nq,Rehman:2009yj}. 
The next important task is to explore the possibility of realizing the gauge symmetry breaking scale $M$ close to a typical GUT scale $\sim 2\times 10^{16}$ GeV. This can, in turn, adequetely suppress the proton decay rate from dimension six operators usually present in GUT models . Achieving $M \sim 2\times 10^{16}$ GeV was one of the main prediction of the original SUSY hybrid inflation model where only radiative correction was included in otherwise a flat potential \cite{Dvali:1994ms}. We, therefore, investigate the possibility of realizing large enough $M$ in SUSY hybrid inflation model with minimal K\"ahler potential, including various important corrections \cite{Dvali:1994ms,Linde:1997sj,Senoguz:2003zw,Rehman:2009nq,Rehman:2009yj}. Specifically, we update the status of SUSY flipped $SU(5)$ hybrid inflation model \cite{Kyae:2005nv,Rehman:2009yj} with minimal K\"ahler potential and soft SUSY masses $\sim 1-100$ TeV.
For other hybrid models of inflation in flipped $SU(5)$ gauge group see \cite{Ellis:2014xda} where each of two hybrid fields is shown to realize inflation. For no-scale SUSY flipped $SU(5)$ models of inflation see \cite{Ellis:2016spb,Ellis:2017jcp}. 

The flipped $SU(5)\equiv SU(5) \times U(1)_X$ model \cite{DeRujula:1980qc,Antoniadis:1987dx} exhibits many remarkable features and constitutes an attractive choice as a grand unified gauge group. In flipped $SU(5)$ model, the doublet-triplet splitting problem is elegantly solved due to the missing partner mechanism \cite{Antoniadis:1987dx}. The proton decay occurs via dimension six operators and is naturally long lived with $M$ around GUT scale. Moreover, it lacks the monopole problem that appears in the spontaneous breaking of other GUT gauge groups (i.e. $SU(5)$, $SU(4)_C \times SU(2)_L \times SU(2)_R$ or $SO(10)$). This property also makes the flipped $SU(5)$ model appropriate choice for the standard version of SUSY hybrid inflation where gauge symmetry is broken after the end of inflation. Finally, flipped $SU(5)$ is also regarded as a natural GUT model due to its connection with F-theory \cite{Jiang:2008yf}.

The outline of the paper is as follows. In the section-2 we briefly introduce the SUSY hybrid model of flipped $SU(5)$ which was first proposed in \cite{Kyae:2005nv}. We update the status of this model with minimal K\"ahler potential in section-3 and check its compatibility with the proton lifetime constraint. The minimal model with $\sim 1-100$ TeV scale soft SUSY masses is shown to predict fast proton decay. However, with the help of leading order non-minimal terms in the K\"ahler potential we overcome this problem and the predictions of inflationary parameters are found to be in accordance with the latest Planck data. This is discussed in detail in section-4. The dominant proton decay mode is $p \rightarrow e^+ \pi^0$ with a lifetime estimated to be of order $10^{36}$ years. Finally, we provide a brief summary of our findings in section-5.

%
%
\section{\large{\bf SUSY FSU(5) hybrid inflation}}%
The minimal Higgs sector of Flipped $SU(5) \equiv FSU(5) \equiv SU(5) \times U(1)_X$ consists of a pair of  Higgs superfields ($10_H, \overline{10}_H$), and a second pair of 5-plet Higgs superfields ($5_h, \overline{5}_h$), which are decomposed under the SM gauge group as
\bea
10_H &=& (10,1) = Q_H (3,2,1/6)+ D^c_H(\overline{3},1,1/3) + N^c_H(1,1,0),  \nonumber \\
\overline{10}_H &=& (\overline{10},-1) = \overline{Q}_H (\overline{3},2,-1/6)+ \overline{D}^c_H(3,1,-1/3) +\overline{N}^c_H(1,1,0),  \nonumber \\
5_h &=& (5,-2) = D_h(3,1,-1/3) + H_d (1,2,-1/2),  \nonumber \\
\overline{5}_h &=& (\overline{5},2) = \overline{D}_h(\bar{3},1,1/3) + H_u(1,2,1/2).
\eea
The MSSM matter content and the right handed neutrino reside in the following representations:
\bea
10^F_i &=& (10,1)_i = Q_i(3,2,1/6)+ D^c_i(\overline{3},1,1/3) + N^c_i(1,1,0),  \nonumber \\
\overline{5}^f_i &=& (\overline{5},-3)_i = U^c_i (\bar{3},1,1/3) + L_i(1,2,1/2),    \nonumber \\
\overline{1}^e_i  &=& (1,5)_i = E^c_i (1,1,+1),
\eea
where $N^c$ is the right handed neutrnio superfield. Assuming the following $R$-charge assignment of the superfields
\be
\left( S, 10_H, \overline{10}_H, 5_h, \overline{5}_h, 10_i, \overline{5}_i,1_i \right) \, = \, \left( 1, 0, 0, 1, 1, 0, 0, 0 \right),
\ee 
the superpotential of the model is given by \cite{Kyae:2005nv},
\bea
W &=& \kappa S \left[ 10_H \overline{10}_H - M^2\right] \nonumber \\
  &+& \lambda_1 10_H 10_H 5_h + \lambda_2 \overline{10}_H \overline{10}_H \overline{5}_h \nonumber \\
  &+& y_{ij}^{(d)}  10^F_i 10^F_j 5_h + y_{ij}^{(u,\nu)} 10^F_i \overline{5}^f_j \overline{5}_h + y_{ij}^{(e)} 1^e_i \overline{5}^f_j 5_h ,   \label{superpot} 
\eea
where the scalar component of the gauge singlet superfield $S$ acts as the inflaton. 
The first line in Eq.~\eqref{superpot} is relevant for inflation and is also responsible for the gauge symmetry breaking of $FSU(5)$ into MSSM as the 10-plet Higgs pair attains non-zero vev in the $N^c_H  , \, \overline{N}^c_H$ direction,
\be
\langle 10_H \overline{10}_H \rangle = \langle N^c_H  \overline{N}^c_H \rangle = M^2.
\ee
The second line in Eq.~\eqref{superpot} contains the terms that are involved in the solution of doublet-triplet splitting problem. The $U(1)_R$ symmetry plays a key role here. This symmetry not only eliminates the $S^2$ and $S^3$ terms to realize successful inflation, it also forbids the bilinear term  $5_h \overline{5}_h$ to avoid GUT scale masses of the MSSM Higgs doublets $H_u$ and $H_d$. The MSSM $\mu$ problem is assumed to be solved by the Giudice-Masiero mechanism \cite{Giudice:1988yz}. Finally, the terms in second line of Eq.~\eqref{superpot} mix the color triplets ($D^c_H,\overline{D}^c_H$) and ($D_h,\overline{D}_h$)  to attain GUT scale masses. This then solves the doublet-triplet problem and eliminates dimension-5 proton decay mediated by colored Higgsino exchange. 

The terms in the third line of Eq.~\eqref{superpot} generate the Dirac mass terms for all fermions, where $y_{ij}^{(d)}$, $y_{ij}^{(u,\nu)}$ and $y_{ij}^{(e)}$ denote the corresponding Yukawa couplings. For a discussion of light neutrino masses in this model see \cite{Kyae:2005nv}. Another possibility to realize light neutrino masses by assuming R-breaking at non-renormaizable level is discussed in \cite{Civiletti:2013cra}. As all matter superfields are neutral under $U(1)_R$ symmetry, an additional $Z_2$ symmetry (or matter parity) is assumed \cite{Kyae:2005nv}. This symmetry not only realizes the possibility of LSP as a cold dark matter candidate but also avoids some unwanted terms in the superpotential.

In the D-flat direction, the relevant part of the global SUSY potential may be written as
\be
V = \kappa^2 \left( \vert 10_H \vert^2 - M^2 \right)^2 + 2 \kappa^2 \vert S \vert^2 \vert 10_H \vert^2 .
\ee
Along the inflationary valley ($\vert 10_H \vert = \vert \overline{10}_H \vert = 0$), SUSY is temporarily broken by the vacuum energy density $V_0 = \kappa^2 M^4$, and is restored later at the global minimum ($\vert \langle 10_H \rangle \vert = \vert \langle \overline{10}_H \rangle \vert = M$, $\vert \langle S \rangle \vert = 0$). 
In the inflationary trajectory, the effective contributions of 1-loop radiative correction and soft SUSY breaking terms can be written as 
\bea
\Delta V_{\text{1-loop}} & \simeq &  \frac{\left(\kappa M\right)^4 \mathcal{N}}{8 \pi^2} F(x),  \label{loopcorec}  \\
\Delta V_{\text{Soft}} & \simeq & a \, m_{3/2} \,  \kappa M^3 x  +  M_S^2\, M^2 x^2,  \label{softmassterm}
\eea
with 
\begin{equation}
	F(x)=\frac{1}{4}\left( \left( x^{4}+1\right) \ln \frac{\left( x^{4}-1\right)}{x^{4}}+2x^{2}\ln \frac{x^{2}+1}{x^{2}-1}+2\ln \frac{\kappa ^{2}M^{2}x^{2}}{Q^{2}}-3\right)
\end{equation}
and 
\begin{equation}
	a = 2\left| 2-A\right| \cos [\arg S+\arg (2-A)].
	\label{a}
\end{equation}
Here, $\mathcal{N}=10$ is the dimensionality of the 10-plet Higgs conjugate pair, $Q$ is the renormalization scale and we have defined $x\equiv |S|/M$. The $a$ and $M_S$ are the coefficients of soft SUSY-breaking linear and mass terms for $S$, respectively and $m_{3/2}$ is the gravitino mass.

\section{\large{\bf Minimal K\"ahler potential}}%
In order to include the supergravity (SUGRA) correction we first consider the minimal canonical K\"ahler potential,
\be
K = \vert S \vert^2 + \vert 10_H \vert^2 + \vert \overline{10}_H \vert^2.
\ee 
The F-term SUGRA scalar potential is given by
\begin{equation}
V_{\text{SUGRA}}=e^{K/m_P^{2}}\left(
K_{i\bar{j}}^{-1}D_{z_{i}}WD_{z^{*}_j}W^{*}-3 m_P^{-2}\left| W\right| ^{2}\right),
\label{VF}
\end{equation}
with $z_{i}$ being the bosonic components of the superfields $z%
_{i}\in \{S, 10_H, \overline{10}_H,\cdots\}$, and we have defined
\be
D_{z_{i}}W \equiv \frac{\partial W}{\partial z_{i}}+m_P^{-2}\frac{%
	\partial K}{\partial z_{i}}W , \,\,\,
K_{i\bar{j}} \equiv \frac{\partial ^{2}K}{\partial z_{i}\partial z_{j}^{*}},
\ee
and $D_{z_{i}^{*}}W^{*}=\left( D_{z_{i}}W\right)^{*}.$
Putting all these corrections together, we obtain the following form of inflationary potential
\begin{eqnarray}
V &\simeq& V_{\text{SUGRA}} + \Delta V_{\text{1-loop}} + \Delta V_{\text{Soft}}, \\
&\simeq&
 \kappa ^{2}M^{4}\left( 1 + \left( \frac{M}{m_{P}}\right) ^{4}\frac{x^{4}}{2}+\frac{\kappa ^{2}\mathcal{N}}{8\pi ^{2}}F(x) + a\left(\frac{m_{3/2}\,x}{\kappa\,M}\right) + \left( \frac{M_S\,x}{\kappa\,M}\right)^2\right).
\label{scalarpot}
\end{eqnarray}
The prediction of various inflationary parameters can now be estimated using standard slow-roll definitions described below.

The leading order slow roll parameters are defined as,
\bea
\epsilon = \frac{1}{4}\left( \frac{m_P}{M}\right)^2
\left( \frac{V'}{V}\right)^2, \,\,\,
\eta = \frac{1}{2}\left( \frac{m_P}{M}\right)^2
\left( \frac{V''}{V} \right), \,\,\,
\xi^2 = \frac{1}{4}\left( \frac{m_P}{M}\right)^4
\left( \frac{V' V'''}{V^2}\right), 
\eea
where $m_P = 2.4 \times 10^{18}$ GeV is the reduced Planck mass. In the leading order slow-roll approximation, the scalar spectral index $n_s$, the tensor-to-scalar ratio $r$ and the running of the scalar spectral index $dn_s / d \ln k$ are given by
\bea
n_s &\simeq& 1+2\,\eta-6\,\epsilon,  \\
r &\simeq& 16\,\epsilon, \label{r0} \\
\frac{d n_s}{d\ln k} &\simeq& 16\,\epsilon\,\eta
-24\,\epsilon^2 - 2\,\xi^2 .
\eea
For negligibly small values of $r$ and $\frac{d n_s}{d\ln k}$, the relevant Planck constraint on the scalar spectral index $n_s$ in the base $\Lambda$CDM model is \cite{Ade:2015lrj}
\be
n_s = 0.9677 \pm 0.0060  \,\,\,\,\,\,\, (68\% CL, Planck\,TT+lowP+lensing).
\ee
The amplitude of the primordial spectrum is given by,
\be
A_{s}(k_0) = \frac{1}{24\,\pi^2}
\left. \left( \frac{V/m_P^4}{\epsilon}\right)\right|_{x = x_0},  \label{curv}
\ee
and has been measured by Planck to be $A_{s}= 2.137 \times 10^{-9}$ at $k_0 = 0.05\, \rm{Mpc}^{-1}$ \cite{Ade:2015lrj}.
The last $N_0$ number of e-folds before the end of inflation is,
\bea
N_0 = 2\left( \frac{M}{m_P}\right) ^{2}\int_{x_e}^{x_{0}}\left( \frac{V}{%
	V'}\right) dx,
\eea
where $x_0$ is the field value at the pivot scale $k_0$, and
$x_e$ is the field value at the end of inflation. 
The value of $x_e$ is fixed either by the breakdown of the slow roll approximation, or by a 'waterfall' destabilization occurring at the value $x_c = 1$ if the slow roll approximation holds.

\begin{figure}[t]
	\centering \includegraphics[width=8.1cm]{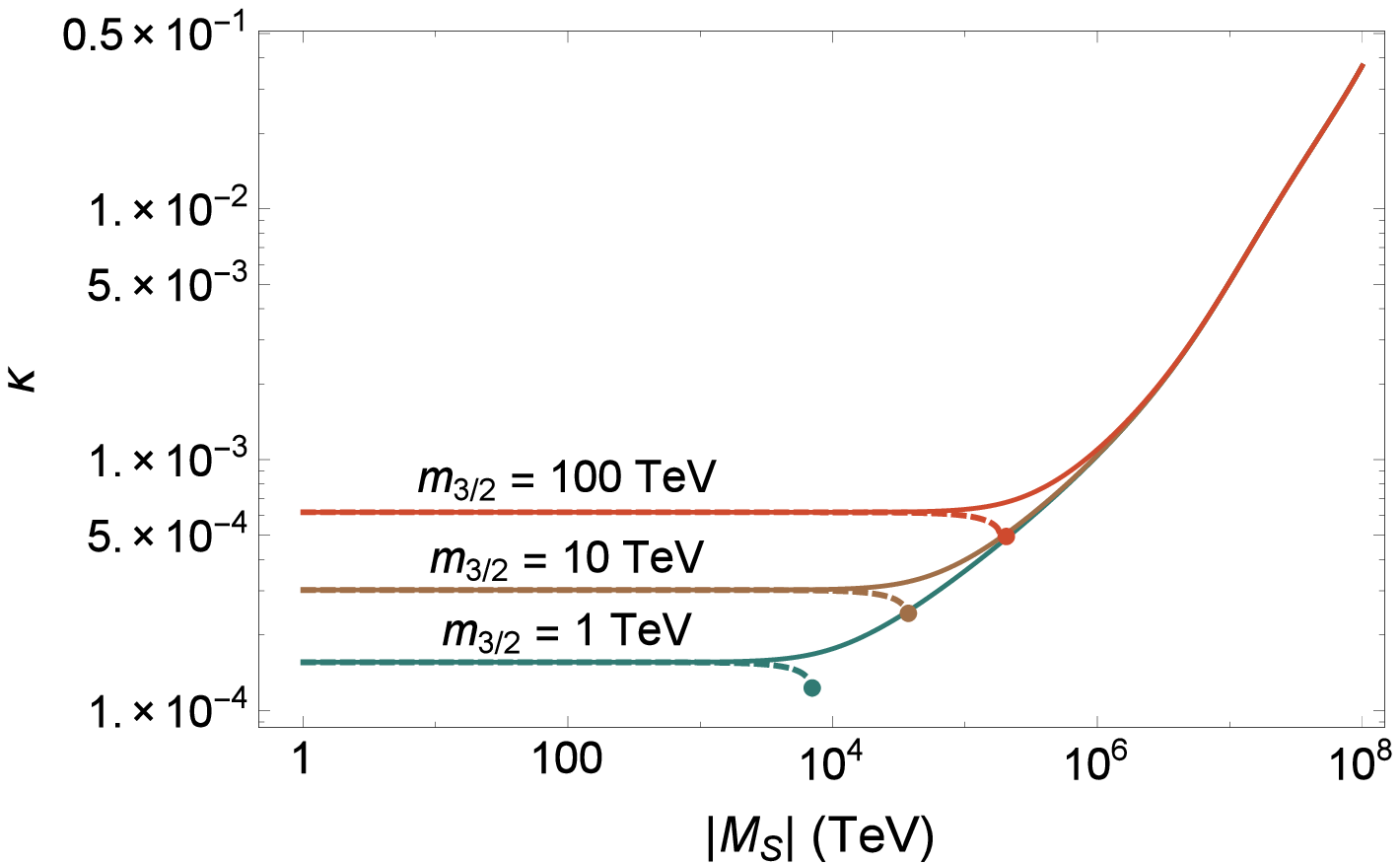} 
	\centering \includegraphics[width=7.9cm]{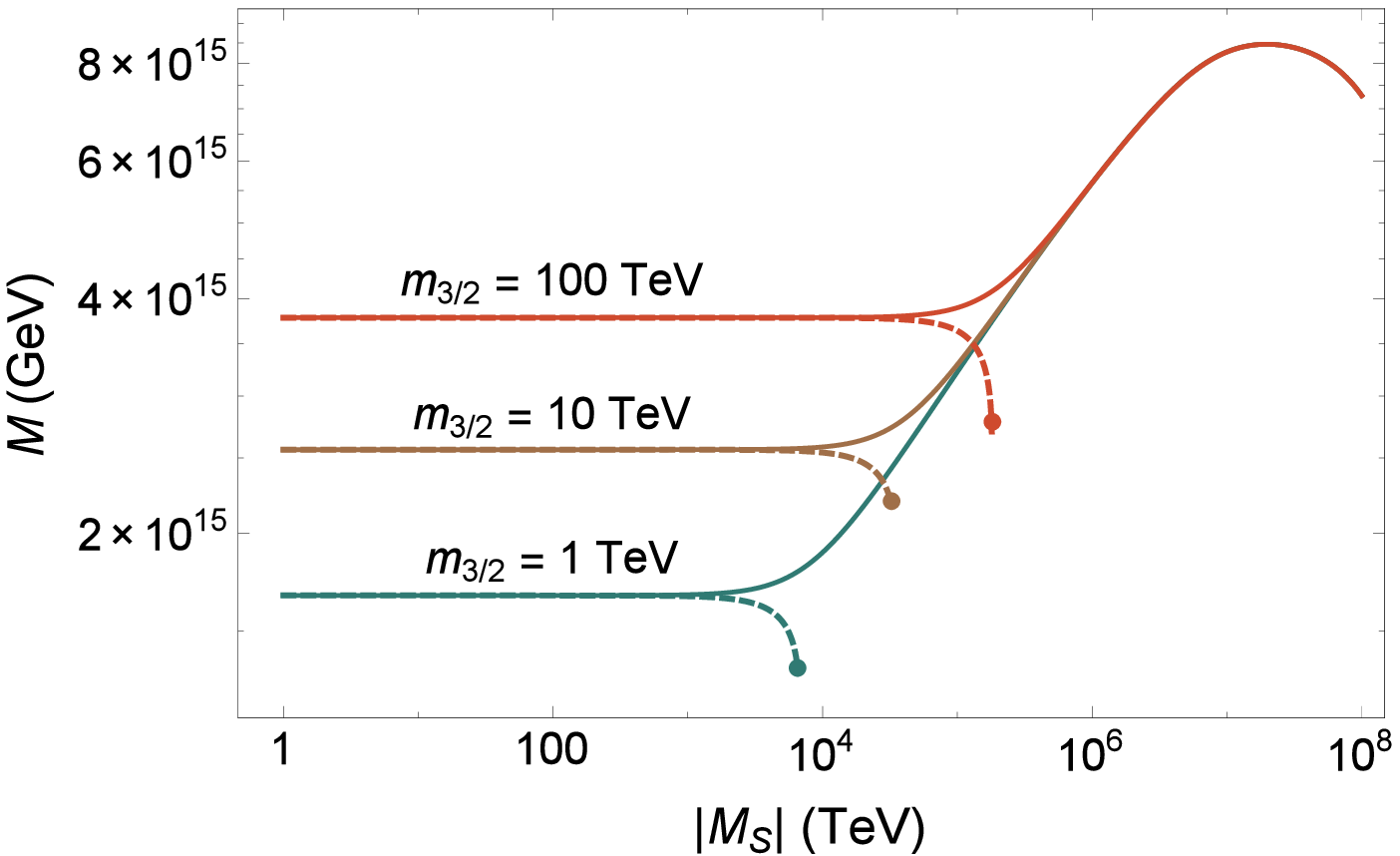}
	\caption{The symmetry breaking scale $M$ (right panel) and $\kappa$ (left panel) versus soft SUSY breaking mass $M_S$ for $a = -1$, $N_0 = 50$ and $n_s = 0.968$ (central value). The green, brown and red curves respectively correspond to $m_{3/2} = 1$, $10$ and $100$ TeV. The solid curves are drawn for $M_S^2 < 0$, while the dashed curves are drawn for $M_S^2 > 0$.}
	\label{fig1}
\end{figure}

\begin{figure}[!t]
	\centering \includegraphics[width=8.25cm]{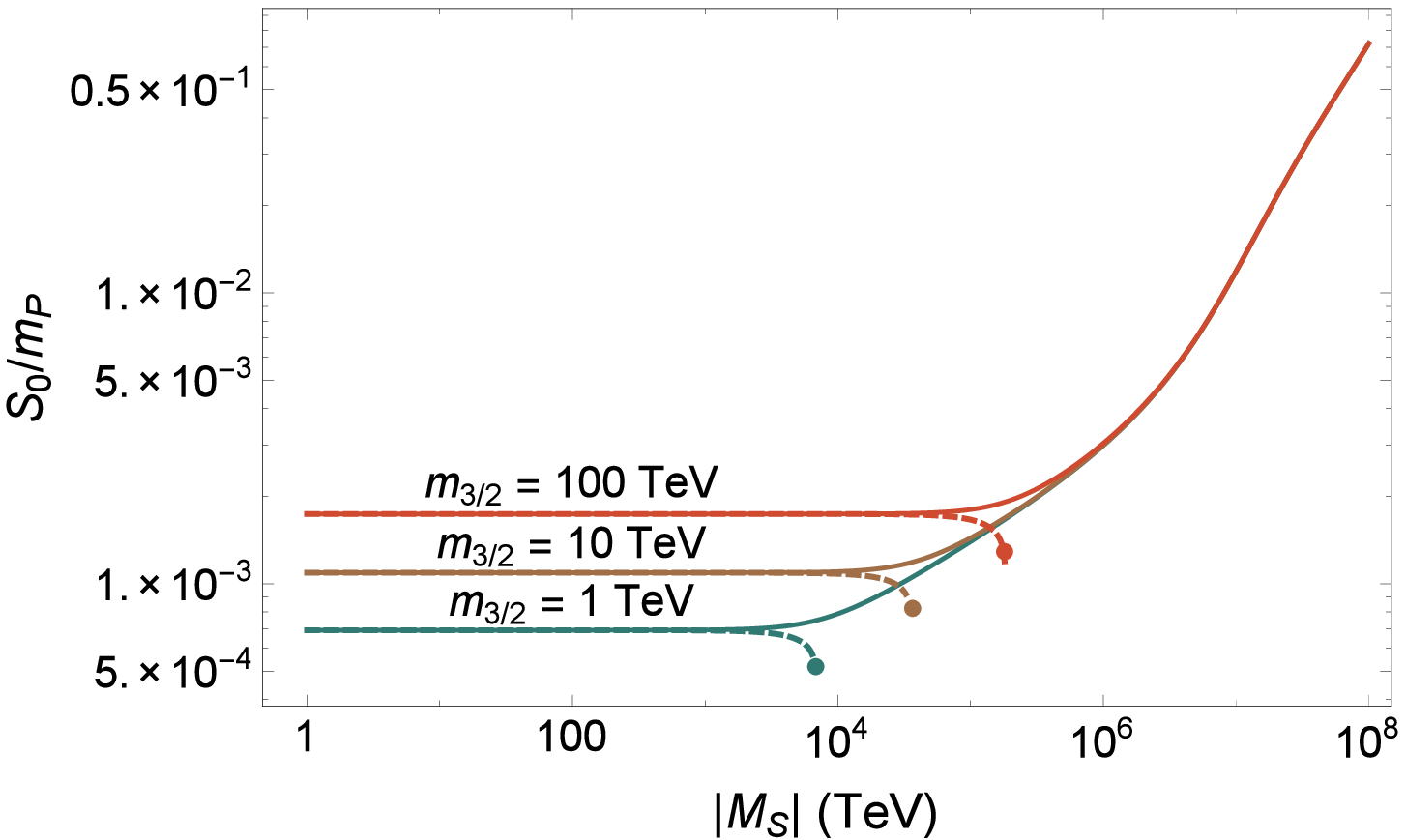}
	\centering \includegraphics[width=7.8cm]{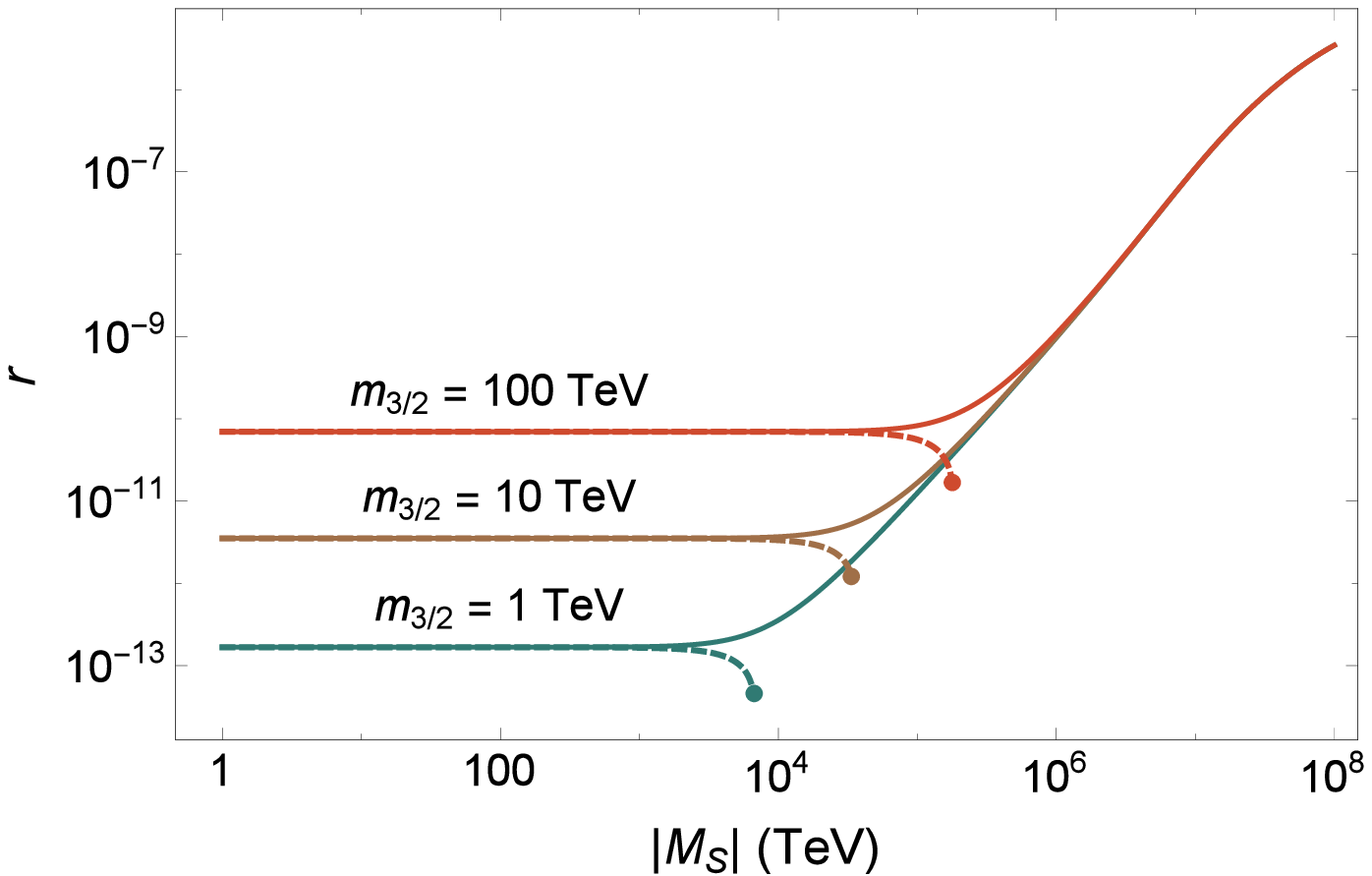} 
	\caption{The tensor to scalar ratio $r$ (right panel) and $S_0/m_P$ (left panel) versus soft SUSY breaking mass $M_S$ for $a = -1$, $N_0 = 50$ and $n_s = 0.968$ (central value). The green, brown and red curves respectively correspond to $m_{3/2} = 1$, $10$ and $100$ TeV. The solid curves are drawn for $M_S^2 < 0$, while the dashed curves are drawn for $M_S^2 > 0$.}
	\label{fig2}
\end{figure}

The results of our numerical calculations are depicted in Figs.~\ref*{fig1} and \ref*{fig2}. Following \cite{Rehman:2009yj}, we have taken $a=-1$ assuming appropriate initial condition for $\arg S$ \cite{urRehman:2006hu}. In addition, we set the number of e-folds $N_0 = 50$ and the scalar spectral index $n_s$ is fixed at the central value ($0.968$) of Planck data bounds. The left panel of Fig.~\ref*{fig1} shows the behavior of $\kappa$ with respect to $M_S$, while the behavior of GUT symmetry breaking scale $M$ with respect to $M_S$ is shown in the right panel and is of particular importance because of proton decay considerations. The curves are drawn for different values of $m_{3/2}$. The solid curves are drawn for $M_S^2 < 0$, while the dashed curves are drawn for $M_S^2 > 0$. 
To make our discussion relevant for the current experiments we restrict the soft masses to be $\sim 1-100$ TeV. This automatically includes the split-susy scenario where soft scalar masses can take values upto 100 TeV \cite{ArkaniHamed:2004fb}. In our region of interest, the radiative correction provides the dominant contribution while the sugra correction is mostly negligible. The suppression of sugra correction is supported by the tiny values of $S_0/m_p \lesssim 2 \times 10^{-3}$ as shown in the left panel of Fig.~\ref{fig2}. This approximation simplifies the expressions of the amplitude of curvature perturbation and the scalar spectral index as
\bea
A_{s} &\simeq& \frac{\kappa^2 M^6}{6 \pi ^2 m_p^6 \left(\frac{\kappa^2 \mathcal{N} F'(x_0)}{8
   \pi ^2} + \frac{2 M_S^2 x_0}{\kappa^2 M^2}- \frac{m_{3/2}}{k M} \right)^2},  \label{curv1} \\
   n_s &\simeq& 1 + \left(\frac{m_p}{M} \right)^2 \left(-\frac{\kappa^2 \mathcal{N} |F''(x_0)|}{8 \pi ^2}+\frac{2
   M_S^2}{\kappa^2 M^2}\right). \label{ns1}
\eea
Next we estimate the values of $\kappa$, $M$ and $M_S$ such that the contributions of the soft linear and mass squared terms are comparable in Eq.~(\ref{curv1}). Assuming $x_0 \sim 1$ we obtain the following expressions using the above equations,
\bea
\kappa &\simeq& \left( \frac{2^{11} \pi ^6  \left(1-n_s\right)}{\mathcal{N}^3 \log
   ^2(4) |F''(x_0)|} \right)^{1/8} \left( \frac{m_{3/2}}{ m_p}  \right)^{1/4},  \label{k} \\
   M &\simeq& \left( \frac{\mathcal{N} |F''(x_0)|^3}{2 \pi ^2 \log
   ^2(4) \left(1-n_s\right){}^3} \right)^{1/8} \left( m_{3/2}m_p^3  \right)^{1/4},  \label{M}   \\
   |M_S|  &\simeq& \frac{\kappa^2 M}{4 \pi } \sqrt{\frac{\mathcal{N}\log (4)}{2}}.  \label{Ms} 
\eea
For $m_{3/2} = 1$ TeV with $F''(x_0)\sim -4.5$ we obtain $\kappa \sim 1.8 \times 10^{-4}$, $M \sim 1.8 \times 10^{15}$ GeV and $M_S \sim 1.1 \times 10^4$ TeV. Similarly, for $m_{3/2} = 100$ TeV with $F''(x_0)\sim -1.7$ we obtain $\kappa \sim 6.5 \times 10^{-4}$, $M \sim 4.1 \times 10^{15}$ GeV and $M_S \sim 2.2 \times 10^4$ TeV. These estimates are  in good agreement with the numerical data shown in Figs.~\ref*{fig1} and \ref*{fig2}. Therefore, with both $m_{3/2}$ and $M_S$ less than 100 TeV, only the radiative correction and the linear soft term dominate while the soft mass squared term and SUGRA corrections are negligibly small. For $M_S \gtrsim 10^4$ TeV, the soft mass squared term begins to take over which drives the curve upward for $M_S^2 < 0$, and downward for $M_S^2 > 0$. Furthermore, the tensor to scalar ratio $r$ turns out to be extremely small, taking on values $r \sim 1.5 \times 10^{-13} - 6.3 \times 10^{-11}$. This can be seen in the right panel of Fig. \ref*{fig2} where $r$ is plotted against $M_S$ for different values of $m_{3/2}$. Our findings in this section are compatible with the results of \cite{Pallis:2013dxa} where SUSY hybrid inflation with minimal K\"ahler and soft SUSY masses of order $\sim (0.1 - 10)$ TeV is considered for $U(1)_{B-L}$ gauge group.
Finally, it is important to note that with a minimal K\"ahler potential and soft SUSY masses of order $\sim (1 - 100)$ TeV, the symmetry breaking scale turns out to be relatively small, $M \sim (1.7 - 3.8) \times 10^{15}$ GeV. This leads to fast proton decay rate as briefly discussed below.

\subsection*{\large{\bf Proton Decay in FSU(5)}}

In the flipped $SU(5)$ model, the dangerous dimension five proton decay operators are highly suppressed due to R-symmetry. For example, even though the operators $10^F_i 10^F_j 10^F_k \overline{5}^f_l \supset Q_i Q_j Q_k L_l$ and $10^F_i \overline{5}^f_j \overline{5}^f_k \overline{1}^e_l \supset D^c_i U^c_j U^c_k E^c_l$ are $FSU(5)$ gauge invariant, they are not invariant under R-symmetry. Further, consider the following R-symmetric $FSU(5)$ gauge invariant operators \cite{Kyae:2005nv},
\be
\frac{S 10^F_i 10^F_j 10^F_k \overline{5}^f_l}{M_P^2}  \supset \frac{\langle S \rangle}{M_P} \left( \frac{Q_i Q_j Q_k L_l}{M_P} \right), \,\,\,\,\,\, 
\frac{S 10^F_i \overline{5}^f_j \overline{5}^f_k \overline{1}^e_l }{M_P^2}  \supset \frac{\langle S \rangle}{M_P} \left( \frac{D^c_i U^c_j U^c_k E^c_l}{M_P} \right).
\ee
With softly broken SUSY, the superfield $S$ is expected to attain non-zero vacuum expectation value $\langle S \rangle \simeq - m_{3/2}/\kappa$ \cite{Dvali:1997uq,King:1997ia}. This makes the above operators heavily suppressed. There are additional R-symmetric $FSU(5)$ gauge invariant operators which lead to proton decay. These include,
\bea
\frac{S 10_H 10^F_i 10^F_j \overline{5}^f_k}{M_P^2}  &\supset& \frac{\langle S \rangle}{M_P}
\frac{\langle N^c_H \rangle}{M_P}  \left( Q_i D^c_j L_k + D^c_i D^c_j U^c_k \right), \\ 
\frac{S 10_H \overline{5}^f_i \overline{5}^f_j \overline{1}^e_k }{M_P^2}  &\supset& \frac{\langle S \rangle}{M_P}\frac{\langle N^c_H \rangle}{M_P} \left( L_i L_j E^c_k \right).
\eea
Although heavily suppressed to have any observable signature for proton decay, these operators are not allowed due to the additional $Z_2$ matter parity imposed on the superpotential (Eq.~(\ref{superpot})). The matter superfields $10^F_i$, $\overline{5}^f_j$ and $\overline{1}^e_l$ are odd under this $Z_2$ matter parity whereas all other superfields are even. This matter parity not only forbids many unwanted couplings as mentioned in \cite{Kyae:2005nv}, but it also makes the neutral LSP a suitable dark matter candidate.  

Therefore, proton decay occurs via dimension six operator from the superheavy gauge boson exchange, and the lifetime for the channel $p \rightarrow e^+ \pi^0$ is given by \cite{Ellis:2002vk,Li:2009fq,Li:2010dp},
\be
\tau \left(p \rightarrow e^+ \pi^0\right) \approx \left(\frac{ M_{5}}{10^{16}\text{ GeV}} \right)^4 \times \left(\frac{ 1.005}{g_{5}} \right)^4 \times 10^{35} \text{ years}, \label{dimsixdecay}
\ee
where $M_{5} = g_{5} M$ is the $SU(3)_c \times  SU(2)_L$ unification scale with unified gauge coupling $g_{5}$ evaluated at this scale. Note that the scale $M_{5}$ usually lies below the unification scale of $SU(5)\times U(1)_X$ group where $g_5$, the gauge coupling of $SU(5)$, is unified with $g_X$, the gauge coupling of $U(1)_X$. The Super Kamiokande experiment places a lower bound on proton lifetime of $1.6 \times 10^{34}$ years at $90\; \%$ confidence level for the channel $p \rightarrow e^+ \pi^0$ \cite{Takhistov:2016eqm,Miura:2016krn}. This then translates into a lower bound on $M$:
\be
 M > 6.3 \times 10^{15} \text{ GeV},
\ee
which disagrees with the result $M \sim (1.7 - 3.8) \times 10^{15}$ GeV stated above.
Moreover, the successful breaking of $FSU(5)$ model into MSSM requires $M \sim 1.6 \times 10^{16}$ GeV for $M_{5}\sim 10^{16}$ GeV and $g_{5} \sim 0.7$. These issues can be resolved by employing a non-minimal K\"ahler potential in which case the symmetry breaking scale $M$ can be raised to the desired value.
\section{\large{\bf Non-minimal K\"ahler potential}}
The non-minimal K\"ahler potential may be expanded as,
\bea
K &=& \vert S \vert^2 + \vert 10_H \vert^2 + \vert \overline{10}_H \vert^2 + \frac{\kappa_S}{4} \frac{\vert S \vert^4}{m_P^2} + \frac{\kappa_H}{4} \frac{\vert 10_H \vert^4}{m_P^2} + \frac{\kappa_{\bar{H}}}{4} \frac{\vert \overline{10}_H \vert^4}{m_P^2} \nonumber \\  &+& \kappa_{S H} \frac{\vert S \vert^2 \vert 10_H \vert^2}{m_P^2} + \kappa_{S \bar{H}} \frac{\vert S \vert^2 \vert \overline{10}_H \vert^2}{m_P^2} + \kappa_{H \bar{H}} \frac{\vert 10_H \vert^2 \vert \overline{10}_H \vert^2}{m_P^2} + \frac{\kappa_{SS}}{6} \frac{\vert S \vert^6}{m_P^4} + \cdots . \label{nonminkhlr}
\eea 
Using Eqs.~\eqref{superpot}, \eqref{VF} and \eqref{nonminkhlr} along with the radiative correction in Eq.~\eqref{loopcorec} and soft mass terms in Eq.~\eqref{softmassterm}, we obtain the following inflationary potential,
\be
	V \simeq \kappa^{2}M^{4} \left( 1 - \kappa_S \left( \frac{M}{m_P} \right)^2 x^2 + \gamma_S\left( \frac{M}{m_{P}}\right)^{4}\frac{x^{4}}{2} + \frac{\kappa ^{2}\mathcal{N}}{8\pi ^{2}}F(x) \right. 
	\left. + a\left(\frac{m_{3/2}\,x}{\kappa\,M}\right) + \left( \frac{M_S\,x}{\kappa\,M}\right)^2\right),\label{scalarpot2}
\ee
where $\gamma _{S}=1-\frac{7\kappa _{S}}{2}+2\kappa _{S}^{2}-3\kappa _{SS}$, and we have retained terms up to $\mathcal{O}\; (\left(\vert S \vert / m_P \right)^4)$ from SUGRA corrections (Recall that during inflation both $\vert 10_H \vert$ and $\vert \overline{10}_H \vert$ are zero). We further assume soft masses, $am_{3/2}$ and $M_S$ to be $\sim 1-100$ TeV, with $a$ and $M_S^2$ either positive or negative. 

\begin{figure}[!t]
	\centering \includegraphics[width=7.8cm]{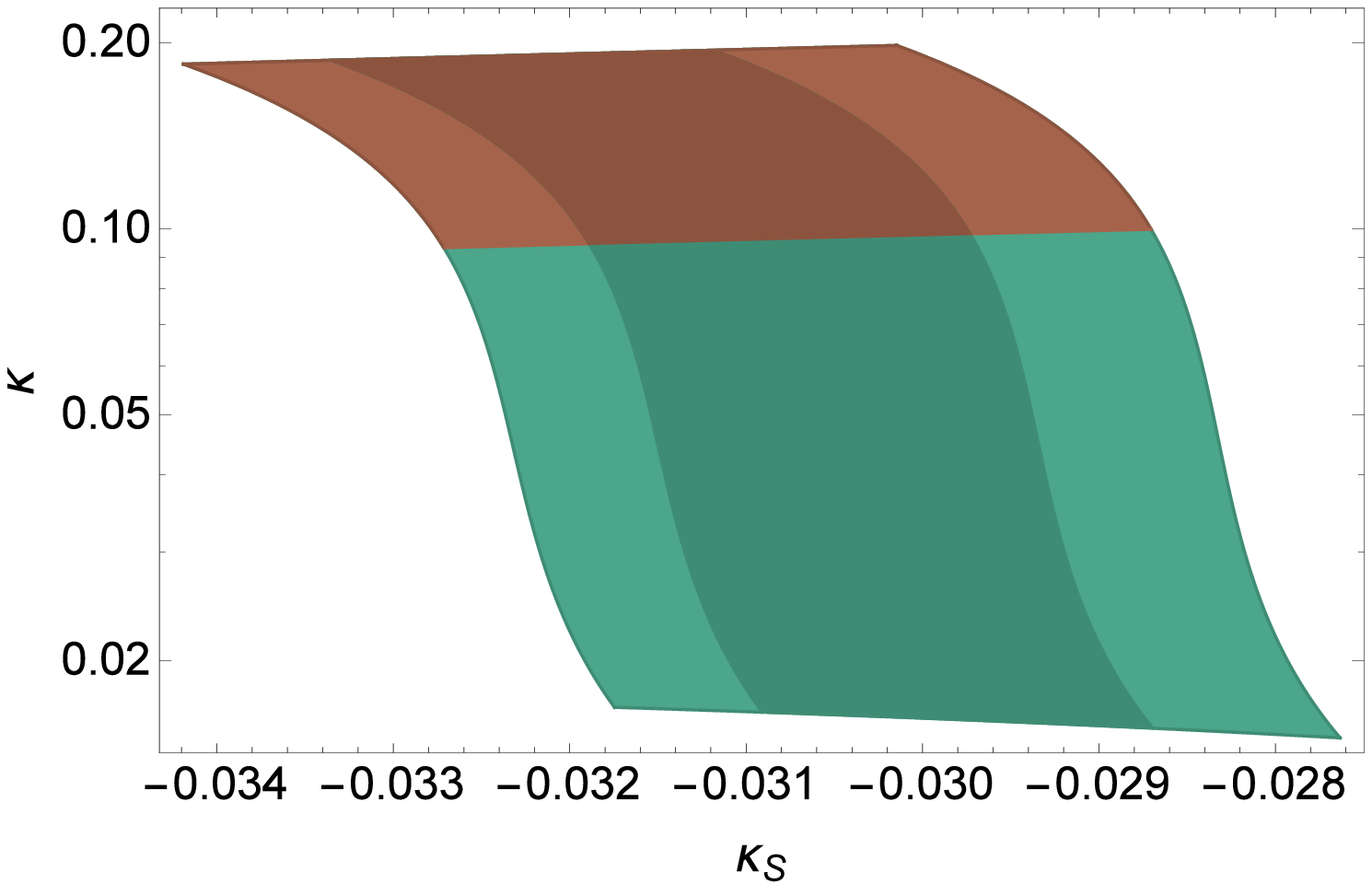} 
	\centering \includegraphics[width=8.15cm]{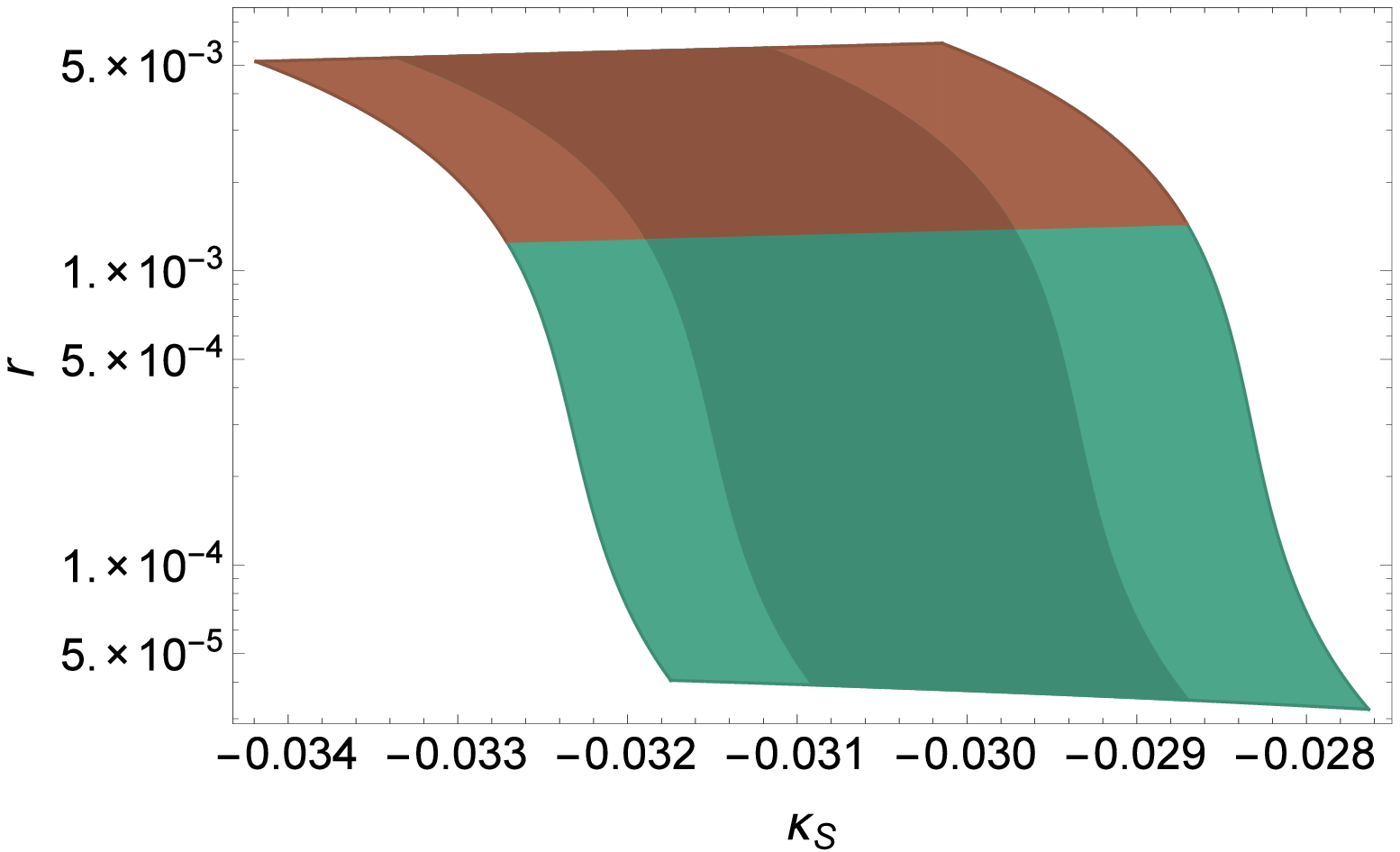}
	\caption{$\kappa$ (left panel) and tensor to scalar ratio $r$ (right panel) with respect to the non-minimal coupling $\kappa_S$ for $N_0 = 50$ and GUT symmetry breaking scale $M \sim 2 \times 10^{16}$ GeV. The lighter shaded region represents the Planck 2-$\sigma$ bounds, while the darker region represents the Planck 1-$\sigma$ bounds. The upper and lower curves corresponds to the $\vert S_0 \vert = m_P$ and $\kappa_{SS} = 1$ constraint, respectively. The brown shaded region represents $\vert S_0 \vert \geq 0.5 \,m_P$.}
	\label{fig3}
\end{figure}

\begin{figure}[t]
	\centering \includegraphics[width=7.8cm]{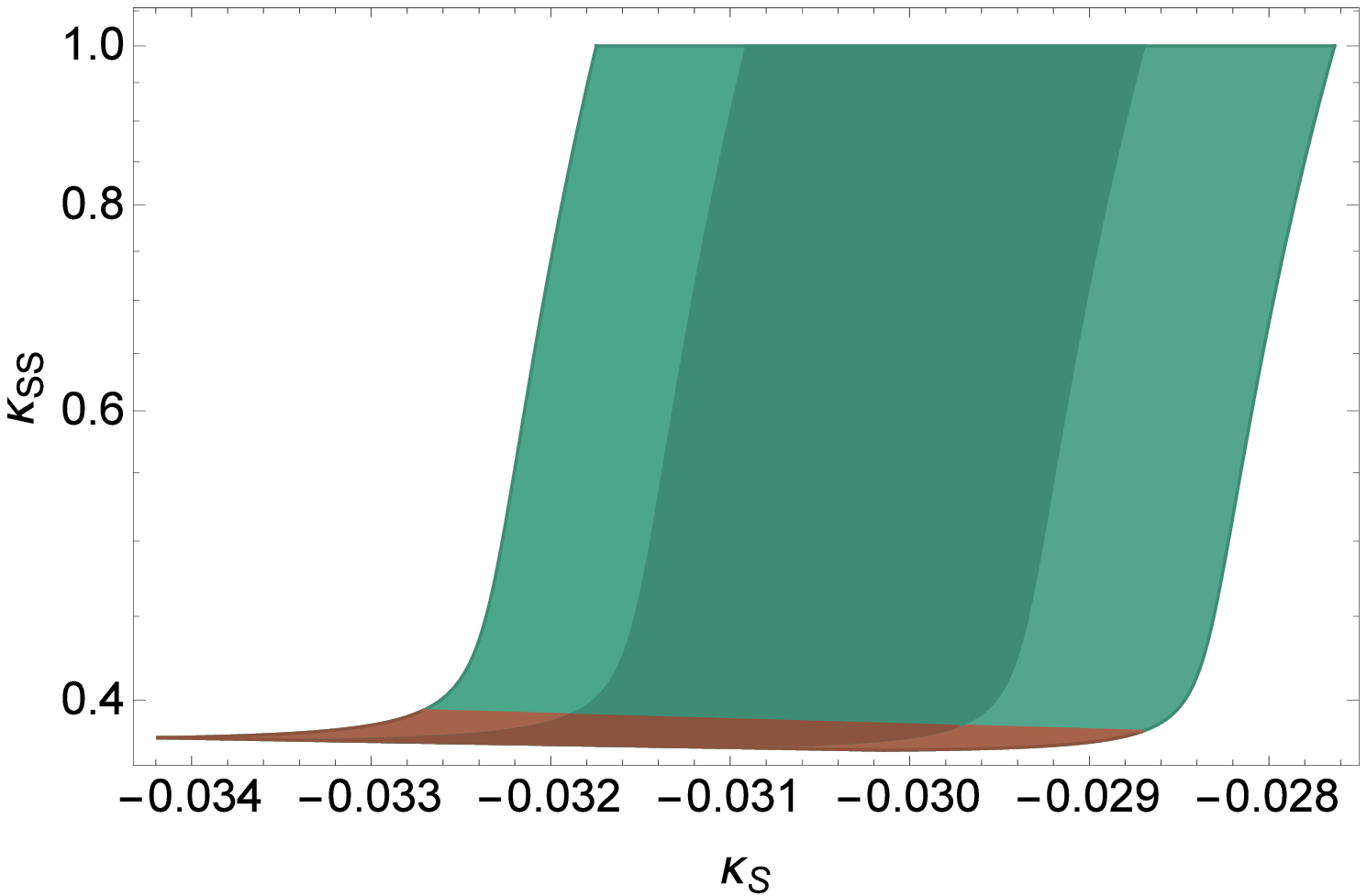} 
	\centering \includegraphics[width=7.9cm]{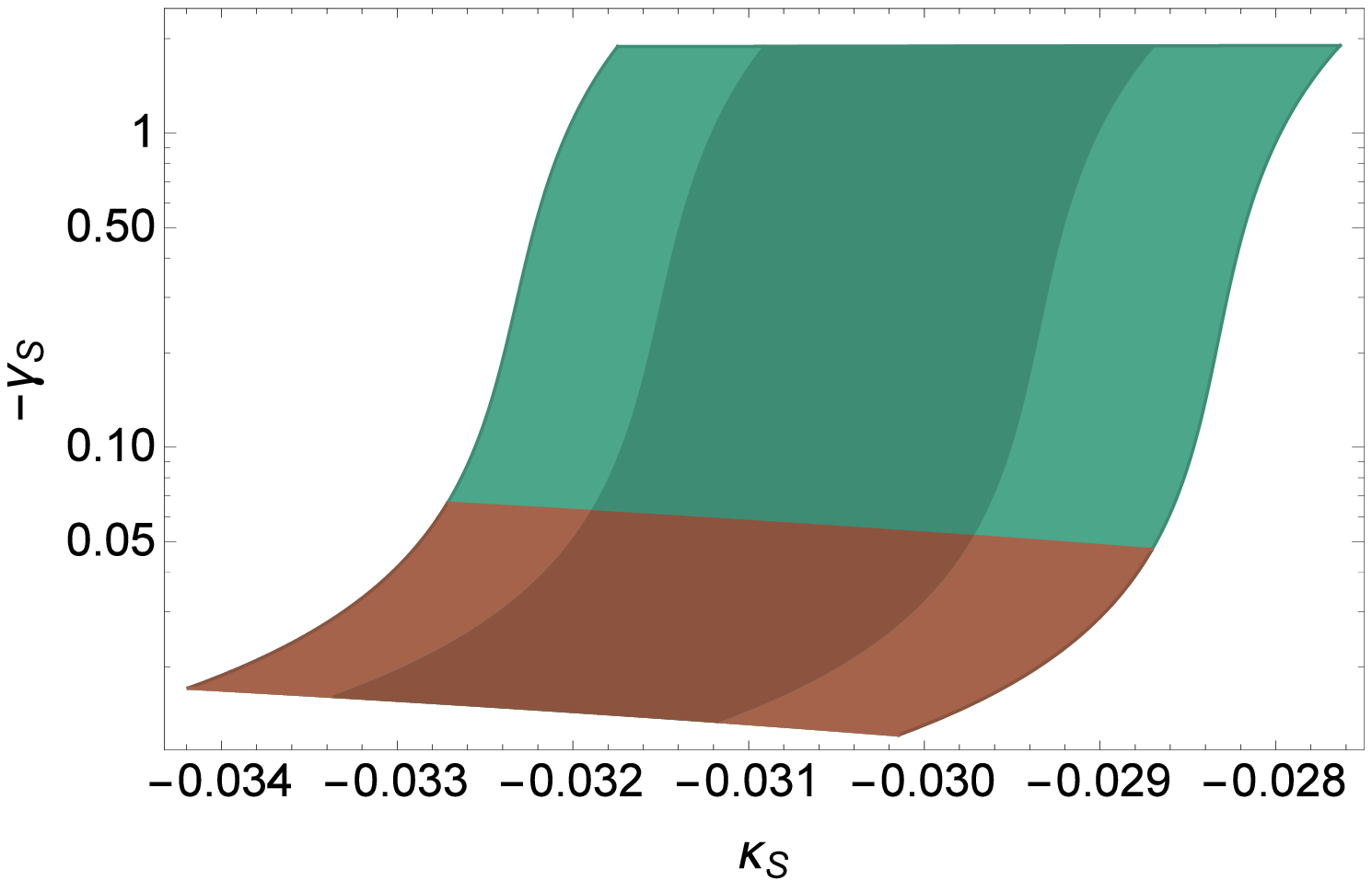}
	\caption{$\kappa_{SS}$ (left panel) and $\gamma_{S}$ (right panel) with respect to the non-minimal coupling $\kappa_S$ for $N_0 = 50$ and GUT symmetry breaking scale $M \sim 2 \times 10^{16}$ GeV. The lighter shaded region represents the Planck 2-$\sigma$ bounds, while the darker region represents the Planck 1-$\sigma$ bounds. The upper and lower curves correspond to the $\kappa_{SS} = 1$ and $\vert S_0 \vert = m_P$ constraints, respectively. The brown shaded region represents $\vert S_0 \vert \geq 0.5\,m_P$.} \label{fig4}
\end{figure}

\begin{figure}[t]
	\centering \includegraphics[width=8.0cm]{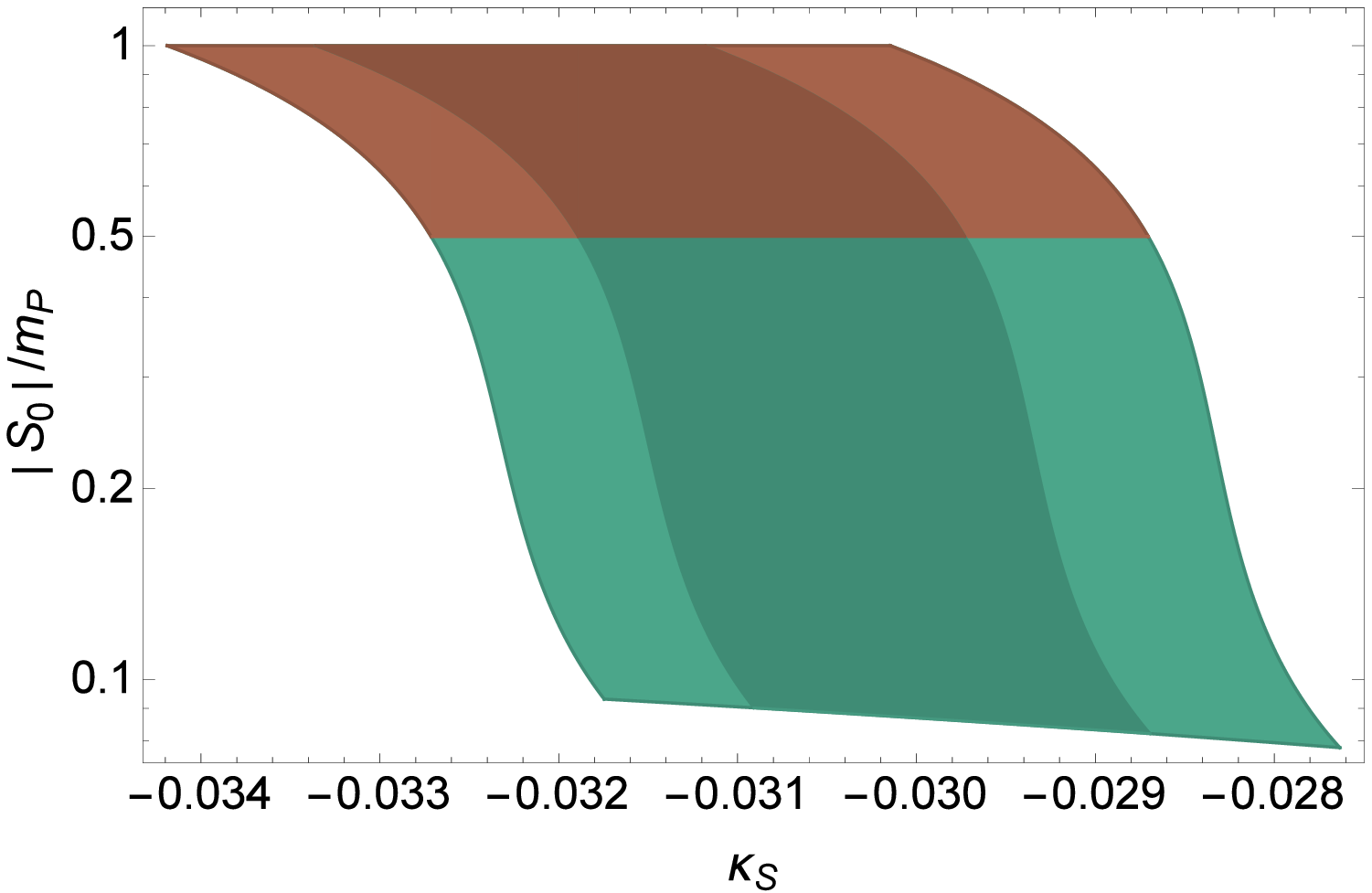} 
	\centering \includegraphics[width=8.0cm]{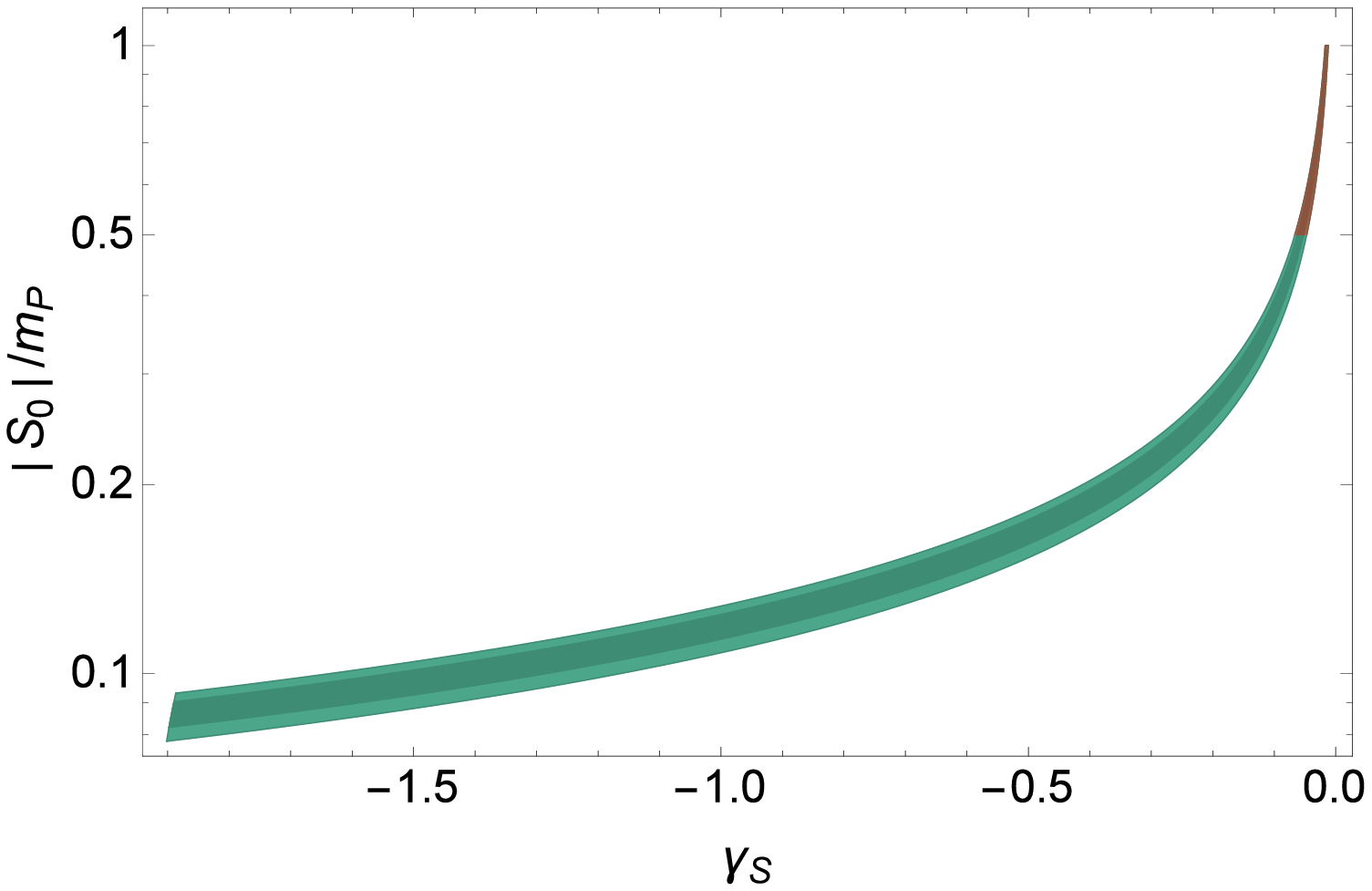}
	\caption{$\vert S_0 \vert / m_P$ versus non-minimal coupling $\kappa_S$ (left panel) and $\gamma_S$ (right panel) for $N_0 = 50$ and GUT symmetry breaking scale $M \sim 2 \times 10^{16}$ GeV. The lighter shaded region represents the Planck 2-$\sigma$ bounds, while the darker region represents the Planck 1-$\sigma$ bounds. The upper and lower curves correspond to the $\vert S_0 \vert = m_P$ and $\kappa_{SS} = 1$ constraints, respectively. The brown shaded region represents $\vert S_0 \vert \geq 0.5 m_P$. The brown shaded region represents $\vert S_0 \vert \geq 0.5\,m_P$.} \label{fig5}
\end{figure}

\begin{figure}[!t]
	\centering \includegraphics[width=7.9cm]{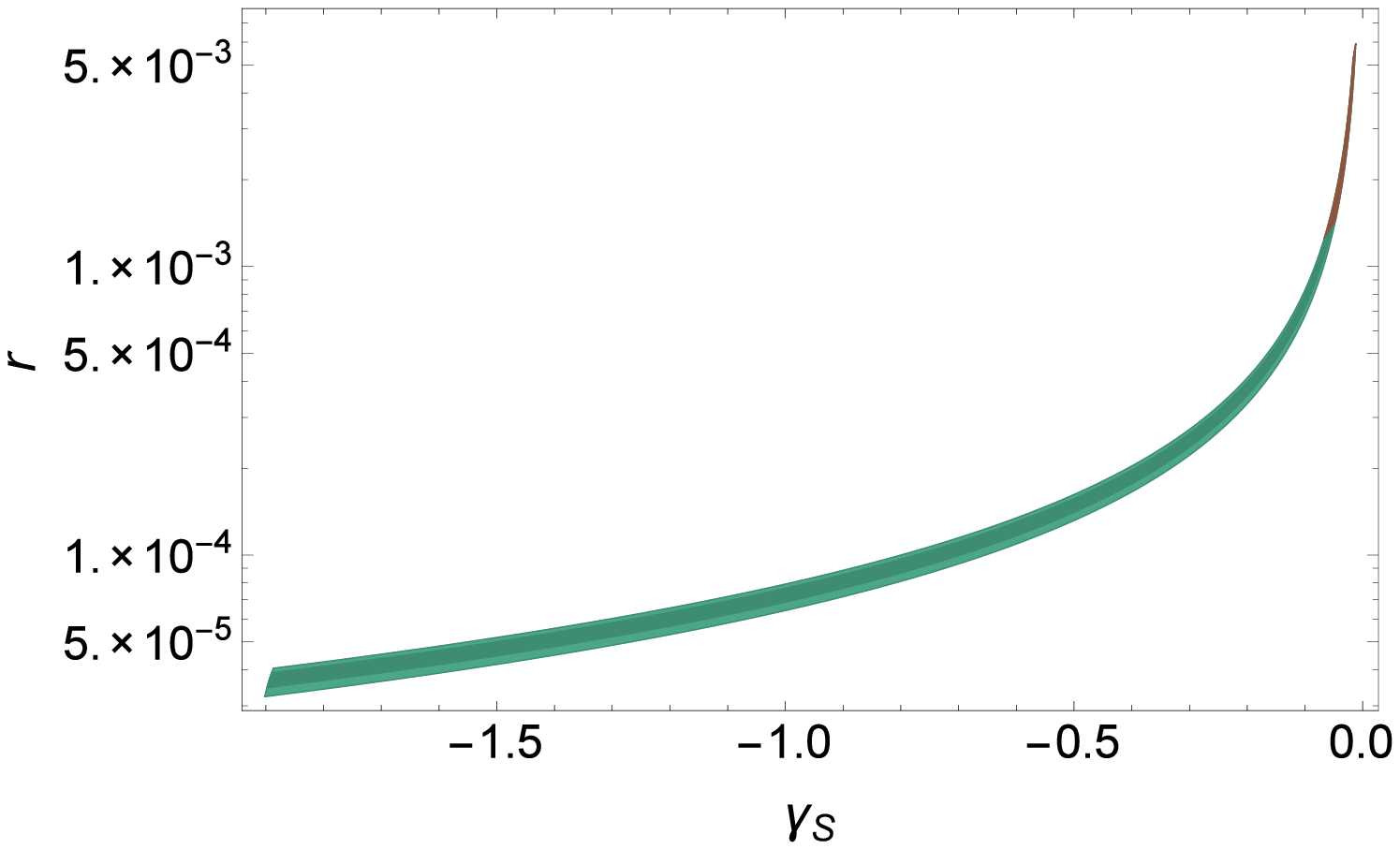} 
	\centering \includegraphics[width=8.1cm]{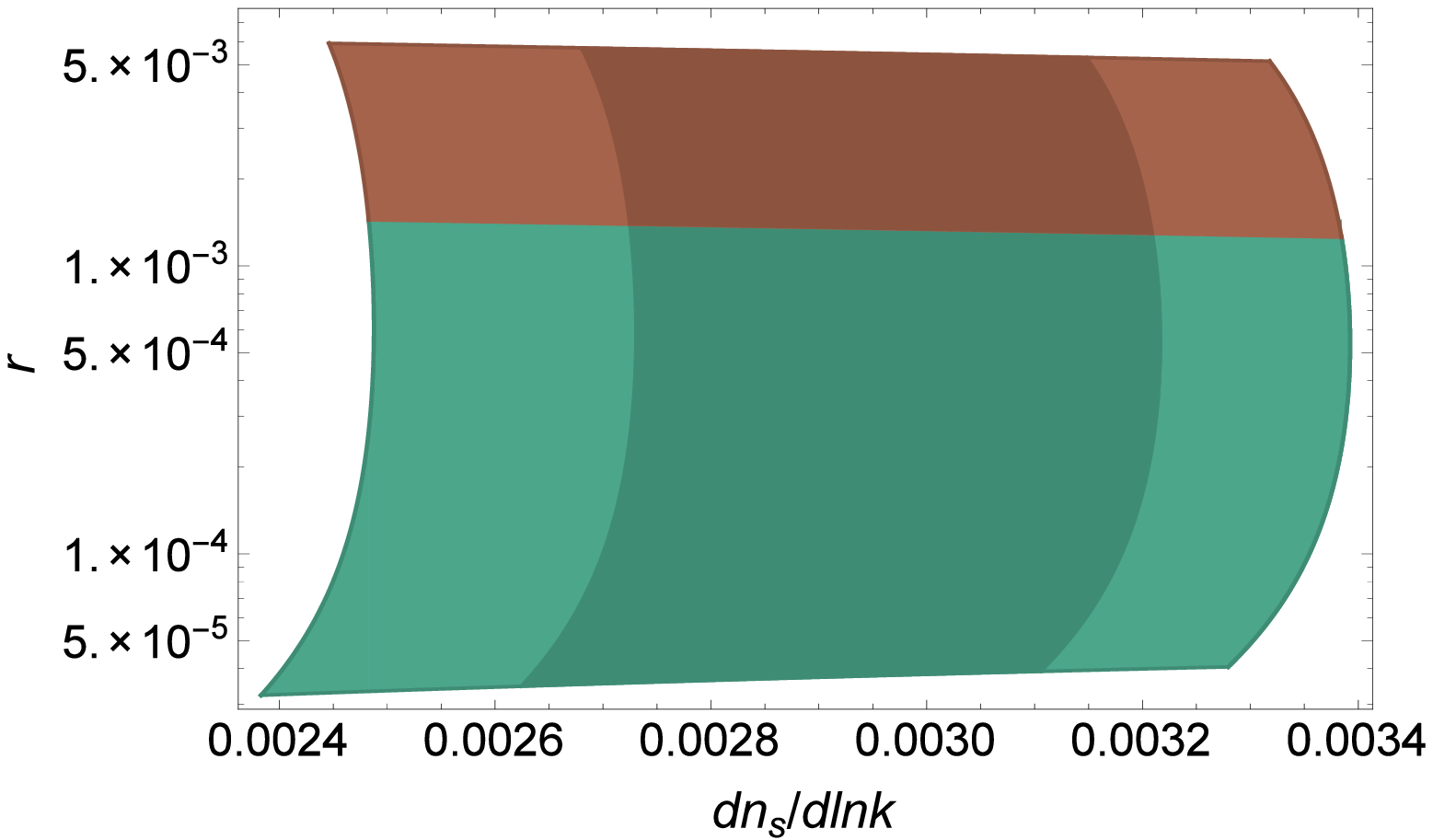}
	\caption{Tensor to scalar ratio $r$ versus the quartic coupling $\gamma_S$ (left panel) and the running of scalar spectral index $dn_s/d \ln k$ (right panel) for $N_0 = 50$, $a = 1$, $m_{3/2} = M_S \sim 1$ TeV ($M_S^2 > 0$) and GUT symmetry breaking scale $M \sim 2 \times 10^{16}$ GeV. The lighter shaded region represents the Planck 2-$\sigma$ bounds, while the darker region represents the Planck 1-$\sigma$ bounds. The upper and lower curves correspond to the $\vert S_0 \vert = m_P$ and $\kappa_{SS} = 1$ constraints, respectively. The brown shaded region represents $\vert S_0 \vert \geq 0.5\,m_P$.} \label{fig6}
\end{figure}

The parameter space consistent with the Planck data bounds is enlarged with the addition of two non-minimal parameters $\kappa_S$ and $\kappa_{SS}$. However, to make our discussion interesting for near future experiments \cite{Andre:2013afa,Matsumura:2013aja}, we restrict ourselves to the parameter region with the largest possible values of the tensor to scalar ratio $r$ with $M \sim 2 \times 10^{16}$ GeV. As previously discussed in \cite{Rehman:2010wm}, the possibility of larger $r$ solutions restricts the non-minimal parameters, namely $\kappa_S < 0$ and $\kappa_{SS} > 0$ with the quartic coupling $\gamma_S<0$. Therefore, large $r$ solutions are obtained mainly with a potential of the form
\begin{equation}
\frac{V}{V_0} \supset 1 + \text{Quadratic} - \text{Quartic}.
\label{potform}
\end{equation}
The linear and soft mass squared terms with $am_{3/2}$ and $M_S $ $\sim 1-100$ TeV are suppressed, while the radiative and SUGRA corrections parametrized by $\kappa_S$ and $\kappa_{SS}$ play the dominant role. To keep the SUGRA expansion under control we further limit $S_0 \leq m_P$. We also require that the non-minimal couplings $|\kappa_S| \leq 1$ and $|\kappa_{SS}| \leq 1$. Using next to leading order slow roll approximation, the results of our numerical calculations are displayed in Figs. \ref{fig3} - \ref{fig6}. The lighter (darker) region represents the Planck 2-$\sigma$ (1-$\sigma$) bounds on $r$ and $n_s$. The upper and lower cutoffs correspond to the $S_0 = m_P$ and $\kappa_{SS} = 1$ constraints. We have also included a brown shaded region for $\vert S_0 \vert \gtrsim 0.5\,m_P$. This is the ultraviolet sensitivity region where higher order Planck suppressed terms in the SUGRA expansion become important. Outside this region with  $\vert S_0 \vert \lesssim 0.5\,m_P$ we not only obtain a natural suppression of higher order terms but also ensure the boundness of the potential, a problem arising due to $\gamma_S<0$ \cite{Civiletti:2014bca}.

Fig.~\ref{fig3} depicts the behavior of $\kappa$ and the tensor to scalar ratio $r$ with respect to the non-minimal coupling $\kappa_S$. These plots resemble one another and their behavior can be understood from the following approximate relation between $r$ and $\kappa$ (from Eqs.~(\ref{r0}) and (\ref{curv})),
\begin{equation}
r \simeq \left( \frac{2  \kappa^2}{3 \pi^2 A_s (k_0)} \right) \left( \frac{M}{m_P} \right)^4 .
\end{equation}
This relation shows that larger values of $r$ are expected when $\kappa$ or $M$ is large. Since $M$ is fixed, larger $r$ values should occur for larger $\kappa$ values. For  fixed $M \sim 2 \times 10^{16}$ GeV, the highest value of $r$ ($\sim 5 \times 10^{-3}$) obtained in our numerical results occurs for $\kappa \simeq 0.2$ (see Fig.~\ref{fig3}). 
In the leading order slow roll approximation, the spectral index $n_s$ and tensor to scalar ratio $r$ are given by
\begin{eqnarray}
n_s &\simeq& 1 - 2 \kappa_S + 6 \gamma_S \left(\frac{M}{m_P}\right)^2 x_0^2 + \left( \frac{m_P}{M} \right)^2 \frac{\mathcal{N} \kappa^2 F''(x_0)}{8 \pi^2}, \label{ns} \\
r &\simeq& 4 \left( \frac{m_P}{M} \right)^2 \left[ 2 \gamma_S \left(\frac{M}{m_P}\right)^4 x_0^3 - 2 \kappa_S \left(\frac{M}{m_P}\right)^2 x_0 + \frac{\mathcal{N} \kappa^2 F'(x_0)}{8 \pi^2} \right]^2. \label{r}
\end{eqnarray}
Solving these two equations simultaneously for $S_0 \simeq m_P$, $r \simeq 5 \times 10^{-3}$ and $n_s \simeq 0.968$ (central value) we obtain $\kappa_{S} \simeq -0.03$ and $\gamma_S \simeq -0.014$. Further, with $\vert S_0 \vert \sim 0.5\,m_P$ and $r \sim 10^{-3}$, we get $\kappa \sim 0.09$, $\kappa_{S} \simeq -0.03$ and $\gamma_S \simeq -0.06$. These estimates are in good agreement with our numerical results displayed in Figs.~\ref{fig3}$-$\ref{fig6}. The behavior of $\kappa_{SS}$ and $\gamma_S$ with respect to  $\kappa_S$ is presented in Fig.~\ref{fig4}, while Fig.~\ref{fig5} depicts the behavior of $S_0/m_P$ with respect to $\kappa_S$ and $\gamma_S$. 

To facilitate this discussion further, we have also provided a plot of $r$ versus $\gamma_S$ in the left panel of Fig.~\ref{fig6} where it can be seen that larger values of $r \, (\sim 10^{-3})$ are obtained for smaller values of $\gamma_S \, (\sim -0.06)$. It is important to note that large tensor modes can be obtained for any value of scalar spectral index $n_s$ within Planck bounds. In short, for non-minimal couplings $-0.034 \leq \kappa_S \leq -0.027$ and $0.37 \leq \kappa_{SS} \leq 1$, we obtain the scalar spectral index $n_s$ within the Planck $2-\sigma$ bounds and tensor to scalar ratio $r$ in the range ($3 \times 10^{-5} - 5 \times 10^{-3}$). Moreover, with the symmetry breaking scale fixed at ($\sim 2 \times 10^{16}$ GeV), the proton is naturally stable with a lifetime of $\sim 2 \times 10^{36}$ years.

The right panel of Fig.~\ref{fig6} shows the dependence of the spectral running $dn_s/d\ln k$ on tensor-to-scalar ratio $r$. It can be seen that the spectral running does not vary appreciably with $r$ and takes on roughly the same values for large and small $r$ values. The scalar spectral running $dn_s/d\ln k$ varies in the range ($0.0024 - 0.0034$) and this justifies the use of latest Planck's data of $\Lambda$CDM+tensors with no running for the purpose of comparing the predictions of this model. Finally, smaller $r$ values ($\sim 10^{-5}$) are obtained for $S_0 \lesssim 0.1\,m_P$ and large $\kappa_{SS} \simeq 1$ for which $\gamma_S$ is negative and rather large ($\sim -2$), as depicted in Figs.~\ref{fig4} - \ref{fig6}. Since both the quadratic and quartic couplings ($\kappa_S,\,\gamma_S$) are negative in this region, the form of potential remains the same as in Eq.~\eqref{potform}. For $\kappa_S>0$ and $\gamma_S<0$ with soft masses $\sim 1-100$ TeV, only tiny values of $r \lesssim 10^{-12}$ are obtained as discussed recently in \cite{Rehman:2017gkm}.

\section{\large{\bf Summary}}
To summarize, we have revisited supersymmetric hybrid inflation in the framework of flipped $SU(5)$ model. We have shown that with a minimal K\"ahler potential and soft SUSY masses of order $(1 - 100)$ TeV, this model predicts a symmetry breaking scale $M \sim (2 - 4) \times 10^{15}$ GeV, for the central value $n_s = 0.968$. This value of $M$ is significantly below the GUT unification scale $2 \times 10^{16}$ GeV and leads to proton lifetime $\tau \sim 10^{32-33}$ years as compared to the current lower limit $\tau_{p \rightarrow e^+ \pi^0} \gtrsim 1.6 \times 10^{34}$ years determined by the Super-Kamiokande collaboration. The tensor-to-scalar ratio also turns out to be extremely small, taking on values $r \sim (10^{-13} - 7 \times 10^{-11})$. By employing non-minimal K\"ahler potential, with soft SUSY masses of order $1-100$ TeV, the symmetry breaking scale $M \sim 2 \times 10^{16}$ GeV is easily achieved within the Planck $\sigma$-bounds on $n_s$ and the proton is naturally of order $\sim 10^{36}$ years. Moreover, larger tensor modes with observable values ($\sim 10^{-4} - 10^{-3}$) are obtained with non-minimal couplings $-0.034 \leq \kappa_S \leq -0.027$ and $0.37 \leq \kappa_{SS} \leq 1$.

\section*{Acknowledgments}
This work is partially supported by the DOE grant No. DE-SC0013880 (Q.S.).


\end{document}